# TDOA-TWR based positioning algorithm for UWB localization system

Authors version

Marcin Kolakowski, Vitomir Djaja-Josko
Institute of Radioelectronics and Multimedia Technology
Warsaw University of Technology
Warsaw, Poland
marcin.kolakowski@pw.edu.pl

# TDOA-TWR based positioning algorithm for UWB localization system

Marcin Kolakowski, Vitomir Djaja-Josko
Institute of Radioelectronics and Multimedia Technology
Warsaw University of Technology
Warsaw, Poland
m.kolakowski@stud.elka.pw.edu.pl, v.djaja-josko@ire.pw.edu.pl

*Abstract*— **Ultra-wideband positioning systems intended for indoor applications often work in non-line of sight conditions, which result in insufficient precision and accuracy of derived localizations. One of the possible solutions is the implementation of cooperative positioning techniques. The following paper describes a cooperative ultra-wideband positioning system which calculates tag position from TDOA and distance between tags measurements. In the paper positioning system architecture is described and an exemplary transmission scheme for cooperative systems is presented. Considered localization system utilizes an Extended Kalman Filter based algorithm. The algorithm was investigated with simulations and experiments. Conducted experiment consisted in fusing results gathered from typical TDOA positioning system infrastructure and ranging results obtained with ultra-wideband radio modules. The research has shown that the use presented cooperative algorithm increases positioning precision.**

## I. INTRODUCTION

Most of ultra-wideband positioning systems are intended for applications inside closed spaces and buildings. There they are challenged by difficult multipath propagation environment, where NLOS (Non Line of Sight) conditions are common and usually unavoidable. These harsh conditions have a negative impact on positioning accuracy and system reliability [4]. One of the possible solutions to above problems are data fusion techniques aggregating data from different systems or sensors. In multi-tag positioning systems additional data can be obtained through exchanging information between tags and evaluating their relative placement. Such approach is called cooperative positioning [1].

In recent years cooperative positioning was considered for applications in different positioning systems. It has been analyzed for GNSS systems to improve positioning accuracy [2] and improve reliability in urban canyon situations [3]. Cooperative positioning was also considered for UWB applications. The study [5] has shown that the use of cooperative techniques can lead to performance improvement but can also lower positioning accuracy for some tag configurations so algorithms taking the geometric configuration of tags and nodes into account are needed.

Implemented cooperative algorithms utilize different sorts of data. One of the most popular data used in cooperative positioning are ranging measurements [6] but algorithms using other types e.g. AoA (Angle of Arrival) can also be found [7]. In this paper a cooperative algorithm basing on Extended Kalman Filter is proposed and evaluated. It is considered for implementation in the UWB system described in the paper.

The presented algorithm uses TDOA (Time Difference of Arrival) measurement results obtained by system infrastructure and results of ranging performed between system tags. Additionally ultra-wideband cooperative positioning system architecture is proposed.

The structure of the paper is as follows. Section II describes the proposed system architecture used by cooperative UWB positioning system. In section III a possible transmission scheme for such system is presented. Section IV describes the positioning algorithm utilizing TDOA and ranging measurements. The results of simulation and experimental algorithm tests are presented in sections V and VI respectively.

## II. SYSTEM ARCHITECTURE

System architecture is presented in Fig.1. It consists of the infrastructure and a set of tags. Infrastructure comprises synchronized anchor nodes, which act as UWB receivers and Wi-Fi access point, which provides a link between anchors and the system controller, where tags positions are calculated and presented.

In the vast majority of positioning systems tags act solely as transmitters. In the proposed solution their capabilities have been extended by introducing functionality of two-way ranging (TWR) between the tags. It allows to gather additional data which can be further used for the position calculation.

Tags send TWR results to the anchors over the UWB interface. Anchors measure time of packet arrivals and transmit all gathered results to the system controller, which implements novel positioning algorithm performing calculations based on both TDOA and TWR ranging results.

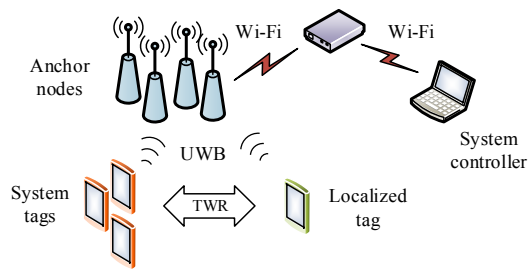

Figure 1. Proposed system architecture

## III. TRANSMISSION SCHEME

Architecture of the cooperative systems should be flexible and easily extendable. Therefore a transmission scheme which minimizes the amount of exchanged packets is needed. The system concept assumes that distance measurements between tags consisting in exchange of messages give anchors an opportunity to carry out TDOA measurements. Both parameters: TDOAs and distances are measured using UWB interface.

The system periodically calculates tags positions. In the proposed transmission scheme, during positioning period each tag sends only one packet. An example of transmission scheme for three tags is presented in Fig.2. The transmission begins with Tag 1 sending a packet. This packet initiates two-way ranging procedure with another tags. Tag 2 answers for the received packet after a fixed delay time $\Delta t_2$. Tag 3 is not triggered by receiving packet from Tag 1, it waits for the packet from Tag 2 instead.

The propagation time between tags is calculated from measured $t_{12}$, $t_{13}$, $t_{23}$ and $t_{3w}$ times and known $\Delta t_2$ and $\Delta t_3$ delay values (1-3).

$$tp_{12} = (t_{12} - \Delta t_2)/2 \quad (1)$$

$$tp_{23} = (t_{23} - \Delta t_3)/2 \quad (2)$$

$$tp_{13} = (t_{13} - \Delta t_3 - t_{3w})/2 \quad (3)$$

Measured values are placed in packets 1 ($t_{12}$ and $t_{13}$), 2 ($t_{23}$) and 3 ($t_{3w}$) and relayed to the system controller by system anchors during next positioning period. The proposed TWR technique suffers from clock frequency errors in system tags. However this issue can be resolved by installing more stable tags oscillators e.g. TCXO (Temperature Compensated Crystal Oscillator) or implementing more advanced ranging methods e.g. SDS-TWR (Symmetrical Double-Sided TWR) [12].

The number of tags used in cooperative systems is usually large so the probability of random events such as tag failure occurring is relatively high. The transmission scheme takes it into account by introducing mechanism of automatic reply for received packets. For example if Tag 2 would be broken and no additional algorithms were implemented Tag 3 would not respond for Tag 1 packet since it wouldn't be triggered by packet 2. In the suggested transmission scheme in case of tag 2 failure it would respond automatically after a fixed delay.

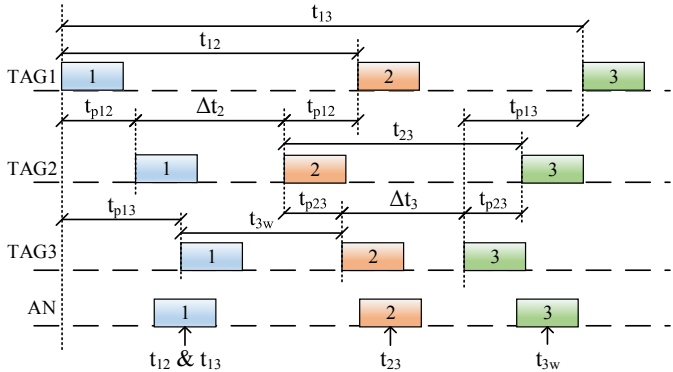

Figure 2. Proposed transmission scheme

## IV. POSITIONING ALGORITHM

The problem of positioning a single tag can be modelled as a dynamic system which state is characterized by actual object location and velocity. Since the tag is positioned on 2D plane the state vector of modelled system consists of x, y coordinates of tag position and x, y velocity vector components (4).

$$x_k = [x \ y \ v_x \ v_y]^T \quad (4)$$

The state transition model in the modelled system is linear, however the measurement model is comprised of non-linear equations. Therefore the positioning algorithm should use an estimator which can handle non-linear models. In the proposed algorithm the Extended Kalman Filter was chosen to estimate system state and process measurement results. The implemented EKF is described by the following set of equations (5-9) [8],

$$\hat{x}_k^- = A\hat{x}_{k-1}^+, \quad (5)$$

$$P_k^- = AP_{k-1}A^T + Q_{k-1}, \quad (6)$$

$$\bar{K}_k = P_k^- H_k^T [H_k P_k^- H_k^T + R_k]^{-1}, \quad (7)$$

$$\hat{x}_k^+ = \hat{x}_k^- + \bar{K}_k(z_k - H_k\hat{x}_k^-), \quad (8)$$

$$P_k^+ = [I - \bar{K}_k H_k]P_k^-, \quad (9)$$

where $\hat{x}_k^-$, $P_k^-$ are a priori estimates of system state and process covariance matrix and $\hat{x}_k^+$ and $P_k^+$ are a posteriori estimates of these values. A is the state transition matrix, $\bar{K}_k$ Kalman gain, Q and R process and measurement noise covariance matrices. Since the measurement model of TDOA is non-linear, measurement sensitivity matrix H is linearized measurement function h(x) (10)

$$h_k(\hat{x}_k^-) = \begin{bmatrix} TDOA_1(\hat{x}_k^-) \\ TDOA_2(\hat{x}_k^-) \\ \vdots \\ TDOA_n(\hat{x}_k^-) \\ d_a(\hat{x}_k^-) \\ d_b(\hat{x}_k^-) \\ \vdots \\ d_m(\hat{x}_k^-) \end{bmatrix} \quad (10)$$

where $TDOA_n(x_k)$ is time difference of arrival calculated for selected anchors, $d_m(x_k)$ is the distance between the tag and tag m. The measurement vector $z_k$ contains TDOA and distance values. Time difference of arrival is calculated from signal arrival times measured by system anchors. Two-way ranging technique measures distances between localized tag and other system tags. Measurement vector length is not fixed. It depends on the number of anchors and tags that are in the range of analyzed tag. If the cooperative option is not used the vector contains TDOA results only. The presented algorithm can be easily extended and adapted for use in three dimensional positioning problem.

## V. SIMULATION RESULTS

The proposed algorithm and simulation environment were implemented using Matlab. The simulator implemented a complete TDOA positioning system. It allowed to simulate positioning process for different number of tags and anchors with or without the use of TWR measurements. The positioning algorithm was tested in a simulated system consisting of 5 anchors and 3 tags. The anchors were placed on the walls of 10x10m square room. There were no obstacles placed in the room – transmission between tags and all anchors proceeded in Line of Sight conditions, so the time measurement results were not biased by transmission delays or multipath propagation.

Simulation tests were conducted for tags placed in randomly selected positions. The locations were chosen from the evenly spaced 0.5m grid placed inside the room. All of the tags were at the same height. Anchors TOA measurement errors were simulated by adding Gaussian noise with standard deviation equal 1 ns. Standard deviation of two-way ranging results was set to 6cm.

The tests were conducted for 500 different tag configurations. For each configuration tags positions were calculated using two versions of algorithm proposed in the paper. In the first case only TDOA results were taken into account. The second version fused both TDOA and TWR measurements results. All of the tests were static, the tags did not change their positions throughout the simulation. The algorithms precision was evaluated by comparing Circular Error Probability (CEP) calculated for 68% of the derived results. The exemplary positioning results are shown in Fig.3. The Cumulative Distributive Function (CDF) curves for CEPs are shown in Fig.4. The CEP values derived for the example presented in Fig. 3 are stored in Table I.

The use of cooperative algorithm improved the quality of the calculated tag positions. Positioning results converged on the real tag location and a significant decrease in CEP was observed. The overall improvement of positioning precision can be assessed basing on a CDF for CEP curve presented in Fig.4. The system fusing TWR and TDOA results is far more precise. The highest CEP value for cooperative system is close to 45 cm. In corresponding TDOA system nearly half of the derived CEP values were worse.

TABLE I. CEP VALUE FOR 68% RESULTS

| point | Used algorithm | |
|---|---|---|
| | *TDOA* | *TDOA & TWR* |
| 1 | 0.37 m | 0.29 m |
| 2 | 0.64 m | 0.29 m |
| 3 | 0.58 m | 0.31 m |

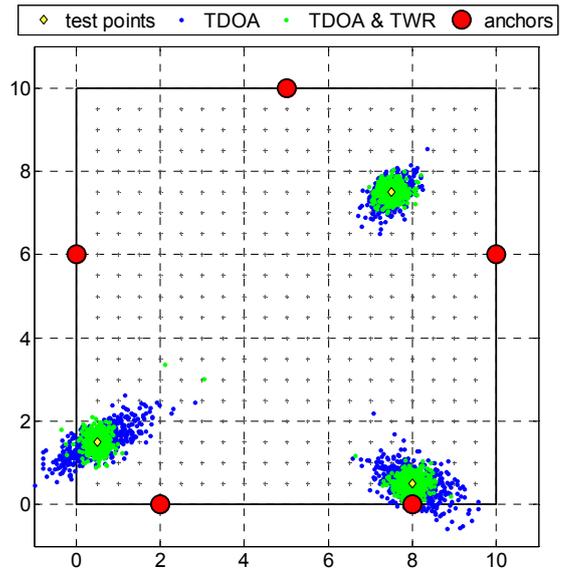

Figure 3. Example of positioning results

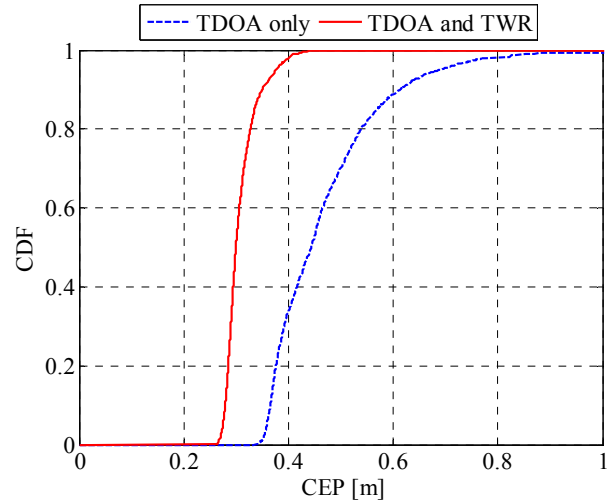

Figure 4. CDF curve for TDOA and TDOA&TWR algorithms

## VI. EXPERIMENTAL RESULTS

Experimental tests were performed in one of laboratories located in the Institute of Radioelectronics and Multimedia Technology. Tests were aimed to check if proposed combined TDOA-TWR architecture would provide better positioning results than standard TDOA technique.

### A. Test setup

The test setup includes UWB positioning system developed within NITICS project [9] and EVK1000 evaluation kit [10]. The system was deployed in the laboratory room of size approx. 6 by 6 meters.

For the TDOA measurements, localization system was used. Infrastructure consisted of six regular anchors and one reference anchor, serving as a source of synchronization signal. All anchors were equipped with DW1000 UWB transceiver [11]: as a part of the DWM1000 module in regular nodes and as a chip with TCXO in the reference node.

For the TWR measurements evaluation kit, EVK1000 was used. It consists of two devices with DW1000 chips. The measurements were taken by DecaRanging application provided with the EVK1000 [12].

System tests were conducted in five points. In each point three hundred TDOA measurements were taken. Additionally the distance between all of the test points was measured by taking three hundred TWR measurements for each pair of points.

Position was calculated using EKF algorithm described in section IV. The performed tests were analogical to those described in section V. From the five points in which the measurements were taken, combinations of three points were chosen. Each point acted as a single tag and their positions were calculated using both versions of the algorithm. Accuracy defined as a mean distance from calculated position to real position and precision defined by CEP68% of the calculated position were taken into account when evaluating the proposed algorithm.

*B. Results*

In Fig. 5 CDF curves for CEP's are shown. As it can be seen, positioning precision has been strongly improved. CDF curves for positioning error are shown in Fig.6. However the use of cooperative algorithm did not improve positioning accuracy. The positioning error became subtly higher. Such effect can be prevented by employing an algorithm selecting the best set of nodes to range with.

Test measurements, involving a functional localization system and TWR data gathering system confirmed simulation results. Overall CEP values are even smaller than in the simulation, since experimental TOA measurements have standard deviations values of roughly 250 ps and TWR standard deviation was about 1.5 cm. Nevertheless, it is proved that novel algorithm, in terms of localization precision, performs better than the standard one.

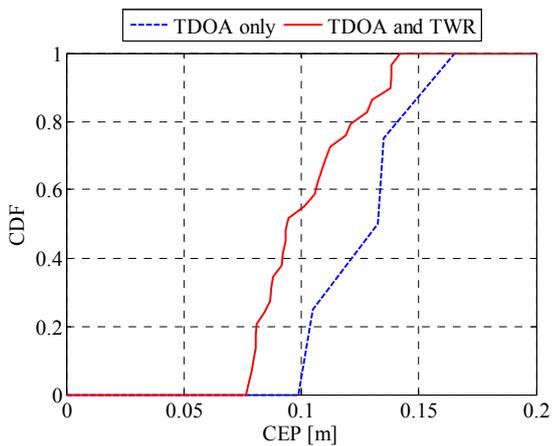

Figure 5. CDF curve for TDOA and TDOA&TWR algorithms

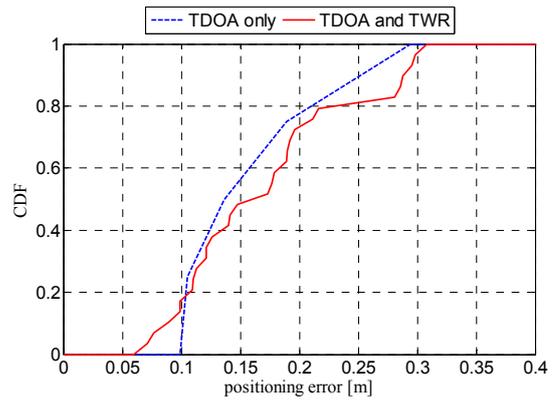

Figure 6. CDF curve for positioning error

VII. CONCLUSIONS

In this paper a positioning algorithm using TDOA and TWR measurements results was presented. The proposed algorithm was evaluated during the simulation end experiment phases. Both results have shown that the use of the cooperative techniques can improve positioning precision. In case of the presented experimental setup, the use of novel algorithm did not result in localization accuracy improvement. In order to achieve such improvement an algorithm choosing cooperating tags is needed.